\documentclass[aps,prl,preprint,superscriptaddress,floatfix]{revtex4-2}
\usepackage[utf8]{inputenc}
\usepackage[T1]{fontenc}
\renewcommand{\selectlanguage}[1]{}
\usepackage{amsmath}
\usepackage{amssymb}
\usepackage{graphicx}
\usepackage{epstopdf}
\usepackage{color}
\usepackage{enumerate}
\usepackage{ulem}
\usepackage{multirow}
\usepackage{array}
\usepackage{units}
\usepackage{float}
\usepackage{amsmath}
\usepackage{hyperref}
\usepackage{epsfig}
\usepackage{bm}
\usepackage{hyphenat}
\usepackage{makecell}
\usepackage{color,soul}
\usepackage[dvipsnames]{xcolor}

\newcolumntype{x}[1]{>{\centering\let\newline\\\arraybackslash\hspace{0pt}}p{#1}}

\usepackage{chemformula}

\begin{document}

\title{Anisotropic magnetoresistance in altermagnetic \ch{MnTe}}

\author{Ruben Dario Gonzalez Betancourt}
\affiliation{Institute of Physics ASCR, v.v.i., Cukrovarnick\'a 10, 162 53, Prague, Czech Republic}
\affiliation{Leibniz Institute for Solid State and Materials Research, IFW Dresden, Helmholtzstr. 20, 01069 Dresden, Germany}
\affiliation{Charles University, Faculty of Mathematics and Physics, Ke Karlovu 3, 121 16 Prague 2, Czech Republic}
\author{Jan Zub\'a\v c}
\affiliation{Institute of Physics ASCR, v.v.i., Cukrovarnick\'a 10, 162 53, Prague, Czech Republic}
\affiliation{Charles University, Faculty of Mathematics and Physics, Ke Karlovu 3, 121 16 Prague 2, Czech Republic}
\author{Kevin Geishendorf}
\affiliation{Institute of Physics ASCR, v.v.i., Cukrovarnick\'a 10, 162 53, Prague, Czech Republic}
\author{Philipp Ritzinger}
\affiliation{Institute of Physics ASCR, v.v.i., Cukrovarnick\'a 10, 162 53, Prague, Czech Republic}
\affiliation{Charles University, Faculty of Mathematics and Physics, Ke Karlovu 3, 121 16 Prague 2, Czech Republic}
\author{Barbora R\r{u}\v{z}i\v{c}kov\'a}
\affiliation{Institute of Physics ASCR, v.v.i., Cukrovarnick\'a 10, 162 53, Prague, Czech Republic}
\author{Tommy Kotte}
\affiliation{Hochfeld-Magnetlabor Dresden (HLD-EMFL), Helmholtz-Zentrum Dresden-Rossendorf, 01328 Dresden, Germany}
\author{Jakub \v{Z}elezn\'y}
\affiliation{Institute of Physics ASCR, v.v.i., Cukrovarnick\'a 10, 162 53, Prague, Czech Republic}
\author{Kamil Olejn\'ik}
\affiliation{Institute of Physics ASCR, v.v.i., Cukrovarnick\'a 10, 162 53, Prague, Czech Republic}
\author{Gunther Springholz}
\affiliation{Institute of Semiconductor and Solid State Physics, Johannes Kepler University Linz, Altenbergerstr. 69, 4040 Linz, Austria}
\author{Bernd B\"uchner}
\affiliation{Leibniz Institute for Solid State and Materials Research, IFW Dresden, Helmholtzstr. 20, 01069 Dresden, Germany}
\affiliation{Institut f\"ur Festk\"orper- und Materialphysik (IFMP), Technical University Dresden, 01062 Dresden, Germany}
\author{Andy Thomas}
\affiliation{Institut f\"ur Festk\"orper- und Materialphysik (IFMP), Technical University Dresden, 01062 Dresden, Germany}
\affiliation{Leibniz Institute for Solid State and Materials Research, IFW Dresden, Helmholtzstr. 20, 01069 Dresden, Germany}
\author{Karel V\'yborn\'y}
\affiliation{Institute of Physics ASCR, v.v.i., Cukrovarnick\'a 10, 162 53, Prague, Czech Republic}
\author{Tomas Jungwirth}
\affiliation{Institute of Physics ASCR, v.v.i., Cukrovarnick\'a 10, 162 53, Prague, Czech Republic}
\affiliation{School of Physics and Astronomy, University of Nottingham, NG7 2RD, Nottingham, United Kingdom}
\author{Helena Reichlov\'a}
\affiliation{Institute of Physics ASCR, v.v.i., Cukrovarnick\'a 10, 162 53, Prague, Czech Republic}
\affiliation{Institut f\"ur Festk\"orper- und Materialphysik (IFMP), Technical University Dresden, 01062 Dresden, Germany}
\author{Dominik Kriegner}
\affiliation{Institute of Physics ASCR, v.v.i., Cukrovarnick\'a 10, 162 53, Prague, Czech Republic}
\affiliation{Institut f\"ur Festk\"orper- und Materialphysik (IFMP), Technical University Dresden, 01062 Dresden, Germany}

\date{\today}

\begin{abstract}

Recently, MnTe was established as an altermagnetic material that hosts spin-polarized electronic bands as well as anomalous transport effects like the anomalous Hall effect. In addition to these effects arising from altermagnetism, MnTe also hosts other magnetoresistance effects. Here, we study the manipulation of the magnetic order by an applied magnetic field and its impact on the electrical resistivity. In particular, we establish which components of anisotropic magnetoresistance are present when the magnetic order is rotated within the hexagonal basal plane. Our experimental results, which are in agreement with our symmetry analysis of the magnetotransport components, showcase the existence of an anisotropic magnetoresistance linked to both the relative orientation of current and magnetic order, as well as crystal and magnetic order.

\end{abstract}

\maketitle

\section{Introduction}

Recently, collinear magnets were classified by spin-group symmetry into three distinct classes. In addition to the conventional ferromagnets and antiferromagnets a third class, altermagnets, was established \cite{smejkal2022, mazin2022}. In this third group of collinear magnets, a compensated magnetic order has opposite magnetic sublattices solely connected by real space rotation possibly combined with translation or inversion transformation \cite{smejkal2022a}. In contrast, conventional ferromagnets have only one spin sublattice or opposite sublattices not connected by any symmetry transformation, and conventional antiferromagnets have opposite sublattices connected by translation or inversion. These distinct symmetry properties reflect in the physical properties of these magnetic materials. In the limit of vanishing relativistic spin orbit coupling, antiferromagnets show spin degenerate bands while ferromagnets and altermagnets can host spin polarized bands. In altermagnets, the spin split band structure is furthermore connected to the rotation symmetry linking the magnetic sublattices. Angle-resolved photoemission experiments identified various altermagnetic compounds, among them MnTe, CrSb, and \ch{RuO2} \cite{krempasky2024, fedchenko2024, lin2024, reimers2024, lee2024}. Hints on altermagnetism in these materials were previously also already obtained by the presence of an anomalous contribution to the transverse magnetoresistance \cite{wasscher1965, reichlova2020, feng2022, tschirner2023, gonzalezbetancourt2023, kluczyk2023} as well as the detection of a spontaneous X-ray magnetic circular dichroism \cite{hariki2023}. In addition to the anomalous transport contributions, altermagnets, as all magnetically ordered materials, host anisotropic magnetoresistance (AMR), i.e., a change of the symmetric components of the resistivity tensor (even under time reversal) upon a directional change of the magnetic order vector \cite{mcguire1975, ritzinger2023}. In MnTe, this was previously used to demonstrate that by field cooling with magnetic field in various orientations, the remanent resistance of the material can be influenced \cite{kriegner2016}. MnTe is a particularly interesting member of the family of the class of altermagnets \cite{mazin2023} since it is a semiconductor with a band gap of approximately 1.4~eV \cite{allen1977, przezdziecka2005, kriegner2016}, and can be doped to host both electron and hole dominated transport \cite{wasscher1969}. Therefore, MnTe enables transport of spin-polarized currents in semiconductors \cite{gonzalez-hernandez2021}. Neutron diffraction and AMR were previously used to determine the magnetic structure and anisotropy of the material \cite{kunitomi1964, przezdziecka2005, szuszkiewicz2006, kriegner2017} (Note that our previous work~\cite{kriegner2016,kriegner2017} concerned the same magnetic structure and we now classify MnTe according to Ref.~\cite{smejkal2022a} as an altermagnet.) In MnTe, the magnetic anisotropy was found to be such that the antiparallel magnetic moments align within the hexagonal basal plane. In undoped MnTe, the magnetic moments can be manipulated in the basal plane by magnetic field \cite{komatsubara1963, kriegner2016, kriegner2017}, suggesting that the anisotropy barrier in the basal plane is significantly smaller than the out-of-plane anisotropy. Additionally, the magnetic anisotropy can be altered by doping \cite{moseley2022} and the influence of the magnetic moment orientation on the electronic structure was studied by first-principle calculations \cite{yin2019, fariajunior2023}.

In this work, we study experimentally how the N\'eel vector can be manipulated by the application of a magnetic field within the basal plane and how the electrical resistivity of the material is changing accordingly. We establish which angle-dependent components of the magnetoresistance anisotropies are expected in MnTe from symmetry analysis. We further experimentally study the angle-dependent variation of the resistivity under the application of magnetic field and extract the various harmonic contributions, and establish their functional dependence on the current and crystal direction. Stoner Wohlfarth modelling \cite{stoner1948} is used to discuss the magnetic field dependence.

\section{Results}

\subsection{Symmetry analysis of the magnetotransport response}

$\alpha$-MnTe crystallizes in the hexagonal NiAs structure (crystallographic space group P$6_3$/mmc \#194, \cite{villars}) depicted in Fig. \ref{Figure1}a. The magnetic moments of the two Mn atoms in the unit cell align antiparallel within the $c$-plane/basal plane of the material with a preferential easy axis alignment along the $[01\bar10]$ or the other two equivalent crystal directions (magnetic space group $Cm^\prime c^\prime m$ \cite{kriegner2017, gonzalezbetancourt2023}). The magnetic anisotropy identified in previous studies \cite{kriegner2017} showed that the magnetocrystalline anisotropy barrier within the basal plane is smaller as compared to the energetic difference for an $c$-axis orientation of the N\'eel vector. We therefore limit our symmetry analysis of the resistivity tensor $\rho_{ij}$ to a rotation of the N\'eel vector within the basal plane. Since our experimental geometry does only allow for a current in the basal plane, we further limit the current direction to be in the basal plane as well. The coordinate system is chosen as $x$ and $y$ along $[2\bar{1}\bar{1}0]$ and $[01\bar10]$, respectively \footnote{Note that we use Bravais indices $hkil$, with $i=-h-k$ for the hexagonal MnTe, while Miller indices $hkl$ are used for the cubic substrate.}. Table \ref{table1} lists the resulting harmonic components of the longitudinal and transverse resistivities in dependence of the current and N\'eel vector orientation, considering terms up to 6-th order in the N\'eel vector angle. The two-fold term in the longitudinal resistivity, depending on the relative orientation of the N\'eel vector with respect to the current, corresponds to the non-crystalline AMR commonly found in \textit{all} magnetically ordered materials, i.e. even in polycrystalline ones. Furthermore, we find that for the longitudinal resistivity, also higher even order terms are allowed due to crystalline symmetry.  In contrast, the transverse resistivity $\rho_{\perp}$ contains in addition to the even AMR (or sometimes called planar Hall effect) terms also odd terms. Here in particular a 3-fold term corresponds to the anomalous Hall effect, i.e. to the antisymmetric off-diagonal component of the resistivity tensor, present due to the altermagnetic nature of the material \cite{gonzalezbetancourt2023, kluczyk2023}.

\begin{table}[h!]
\begin{center}
\begin{tabular}{c x{4cm} x{4cm}} 
 \hline
 $n$-th order & $\rho_{\parallel}$  &  $\rho_{\perp}$ \\ [1ex]
 \hline 
 $2^{nd}$ & $\rho^{(2)} \cos(2(\varphi-\theta))$ & $-\rho^{(2)} \sin{(2(\varphi-\theta))}$ \\ [1ex]
 
 $3^{rd}$ & $0$ & $\rho^{(3)} \sin{(3\theta)}$ \\ [1ex]
 
 $4^{th}$ & $\rho^{(4)} \cos(2\varphi+4\theta)$ & $-\rho^{(4)} \sin{(2\varphi+4\theta)}$ \\ [1ex]
 
 $5^{th}$ & $0$ & $0$ \\ [1ex] 
 
 $6^{th}$ & $\rho^{(6)} \cos(6\theta)$ & $0$ \\[1ex] 
 \hline
 
\end{tabular}
\caption{Symmetry analysis of the longitudinal and transverse resistivity components in MnTe. The angle $\varphi$ is hereby defined as the angle between the $x$ axis ($[2\bar{1}\bar{1}0]$) and the current direction, $\theta$ is the angle between the N\'eel vector and the $x$ axis within the basal plane and $\rho^{(i)}$ represents the amplitude of the respective contribution.}
\label{table1}
\end{center}
\end{table}

\subsection{Sample characterization and fabrication}

\begin{figure*}
        \centering
	\includegraphics[width=1\textwidth]{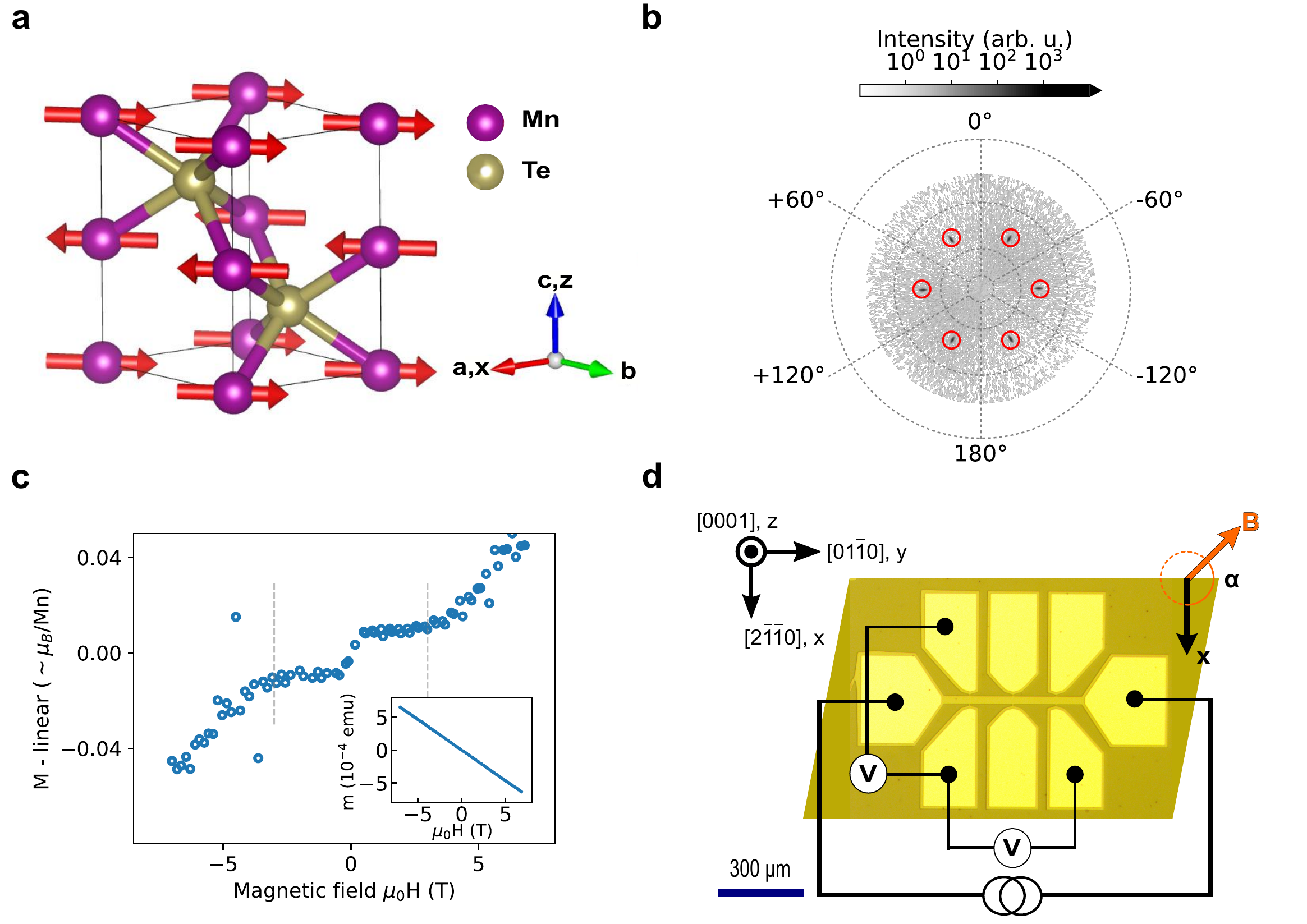}
	\caption{\textbf{MnTe structure, crystallographic and magnetic properties of the thin film as well as device geometry.}
 (a) Depiction of the crystal and easy axis configuration of the $\alpha$-MnTe structure. (b) X-ray diffraction pole figure measurement of the MnTe $1\bar102$ Bragg diffraction shown in stereo-graphic projection. Red circles mark the expected diffraction positions of single crystalline MnTe. (c) SQUID magnetometry data of the MnTe$/$InP sample for in-plane magnetic field at $T=50~K$. To emphasize the lack of residual magnetization, we subtracted a linear slope near zero magnetic field from the data shown in the inset. It is important to note that the subtracted slope encompasses both the susceptibility of MnTe and the diamagnetic substrate. The dashed lines indicate the spin-flop field of $\sim 3$~T.(d) Microscopy image of a lithographically processed Hall bar as well as a sketch of the employed electrical schematic, coordinate system and definition of the magnetic field geometry for angle-dependent resistivity measurements.}
	\label{Figure1}
\end{figure*}

To experimentally test the existence of these terms, we employ magneto transport experiments in single crystalline MnTe thin films grown by molecular epitaxy with a typical thickness of 35~nm. The hexagonal structure of these films is evident from the X-ray diffraction pole figure measurement presented in Fig.~\ref{Figure1}b showing a six-fold symmetric signal at the expected positions. In-plane magnetometry measurements in Fig.~\ref{Figure1}c show the dominant diamagnetic signal of the substrate with no detected spontaneous magnetization consistent with the compensated magnetic structure depicted in panel (a) \cite{gonzalezbetancourt2023, kriegner2017}. After subtraction of a linear contribution, remaining features in the magnetometry data indicate a change of the susceptibility around $\sim3$~T which is related to the reorientation of the magnetic order (spin-flop). Additionally, the very small step-like signal is most probably related to a magnetic impurity which is commonly found in commercial substrates and it has no correspondence to the magnetotransport data.

The films provide the ideal platform to study magnetotransport within the basal plane.
For this purpose we pattern Hall bar structures with various in-plane orientations. Electrical transport is enabled by intrinsic $p$-type conductivity commonly observed in nominally undoped MnTe \cite{wasscher1969, ferrer-roca2000, przezdziecka2005}. 

\subsection{Magneto-transport}

\begin{figure*}
        \centering
	\includegraphics[width=1\textwidth]{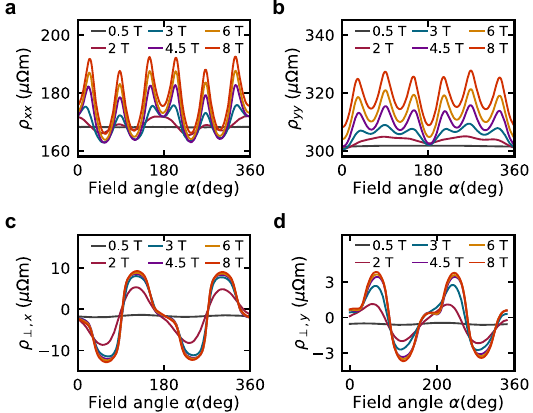}
	\caption{\textbf{In-plane angular-dependent magnetoresistance measurements at $T = 50 K$}. Panels (a) and (c) display longitudinal and transverse resistivity for a Hall bar along the $x$-direction, while panels (b) and (d) represent the same for the $y$-direction. Selected measurements from the probed magnetic field range up to 8~T are shown. Note that here, only the clockwise field rotation data are shown for clarity.}
	\label{Figure2}
\end{figure*}

For our measurements, we mainly employ angular-dependent magnetoresistance (ADMR) measurements, where we rotate an in-plane magnetic field and track the corresponding resistivity change. Note that, this is apriori different from AMR where the dependence of the resistivity on the angle of the magnetic moments is considered (and moreover, symmetry requirements on the response are imposed), which is why we will use these two expressions distinctly in this manuscript. The magnetic field angle $\alpha$ is defined between the $x$-axis, i.e., 90 degrees rotated from one of the magnetic easy axis, and the direction of the applied magnetic field in the sample plane as shown in Fig.~\ref{Figure1}d. Figure~\ref{Figure2} shows ADMR measurements of the longitudinal resistivities $\rho_{xx}$ and $\rho_{yy}$ as well as transverse resistivities $\rho_{\perp,x}$ and $\rho_{\perp,y}$ for current along $x$ and $y$ direction, respectively. While we show only the data for clockwise rotation of the magnetic field in Figure~\ref{Figure2} both rotation data are shown in the supplementary Fig. S1. For strong fields, i.e., 8~T, a clear harmonic variation of the resistivity with the sample angle is observed which is distinct for longitudinal and transversal resistivity as well as for the different current directions.
 
\subsection{Fourier component analysis}

\begin{figure*}
        \centering
	\includegraphics[width=1\textwidth]{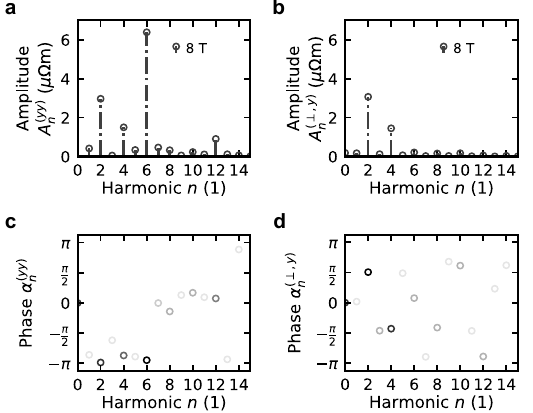}
	\caption{\textbf{Frequency and phase analysis for angular-dependent magnetoresistance}. Panels (a, c) and (b, d) show the amplitudes and phase offsets of the frequency components for the longitudinal resistivity and transverse resistivity signals for current along $y$, respectively. The analysis is shown for the data set recorded at magnetic field of 8~T shown in Fig. \ref{Figure2}b,d.}
	\label{Figure3}
\end{figure*}

To understand our experimental results, we perform a Fourier component analysis, where we separate the experimental data into a sum of $A_n \cos (n\alpha + \alpha_n)$ terms. $A_n$, $\alpha_n$ correspond to the amplitude and phase offset of the $n$-th harmonic, respectively, and should not be confused with $\rho^{(n)}$ in Tab.~\ref{table1} which refer to angle $\theta$ rather than $\alpha$.
We find that in the longitudinal signal the amplitudes of the harmonics of the order 2, 4 and 6 dominate the signal. Additionally, a weaker but clearly detectable signal is present for $n=12$, which will be discussed further below. In the transversal resistivity, we find only clear signals for $n=2$, and 4. The strong contributions with a phase offset near 0 or $\pi$ for the longitudinal resistivity exhibit cosine-like behavior, and the ones with a phase offset near $\pm\pi/2$ for the transverse resistivity exhibit a sine-like behavior. These patterns align with predictions from the symmetry analysis summarized in Table~\ref{table1} which is also consistent with earlier measurements on Corbino devices~\cite{kriegner2017}. The symmetry analysis further implies that, the absolute values of the amplitudes of the two-fold and four-fold components must be equivalent for the longitudinal or transverse resistivity. The sign of the four-fold component, however, should flip between the longitudinal and transverse resistivity. Indeed, this is the case since we find in Fig.~\ref{Figure3}a,b consistent amplitudes for $n=2$, and 4. However, while both components have a phase near $-\pi$ for the longitudinal data, the phase is approximately $\pi/2$ for $n=2$ and $-\pi/2$ for $n=4$ in the transversal data, confirming a difference in sign.

Comparing the component analysis of the ADMR data for currents along $y$ and $x$ directions (c.f. supplementary Fig.~S2) we are able to check the functional dependence of the harmonic components on the current direction. 
In particular, we can identify which terms depend solely on the relative orientation of the magnetic moments and the crystal direction. We find the phase of the six-fold term changing effectively unchanged (change by approximately $2\pi$) indicating that this term depends on the orientation of magnetic moments and the crystal direction consistent with Table~\ref{table1}. In contrast, the phase of the two-fold and four-fold terms is changing by approximately $\pi$, i.e. equal to twice the rotation of the current direction, which is again what we obtain from our symmetry analysis.
Note that here we assume that for sufficient field amplitude the magnetic moments rotate with the field which we verify below.

\begin{figure*}
        \centering
	\includegraphics[width=1\textwidth]{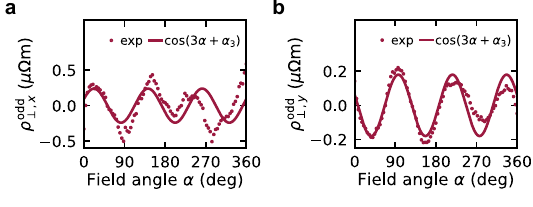}
	\caption{\textbf{Odd-in-magnetic field component of the angular-dependent transversal magnetoresistance due to anomalous Hall effect.} 
    Panels (a, b) show the odd contributions to $\rho_{\perp,x}$ and $\rho_{\perp,y}$ obtained after subtraction of the dominating even-in-field contribution from raw data at 2~T shown in Fig.~\ref{Figure2}c,d. The even-in-field contribution was determined by fitting $\sum_{n=2,4,6,8,10} A_n \cos(n\alpha + \alpha_n)$.
    The odd component is dominated by a three-fold symmetry which is highlighted by the continuous line representing a guide to the eye proportional to $A_{3}\cos(3\alpha + \alpha_3)$.
    }
	\label{Figure4}
\end{figure*}

What makes altermagnets different to $\mathcal{PT}$-symmetric collinear antiferromagnets is the possible presence of the anomalous Hall effect. 
The three-fold component in Tab.~\ref{table1} would be forbidden in $\mathcal{PT}$-symmetric collinear antiferromagnets and its presence, in turn, is characteristic of an altermagnet.
In our experiments, this component is weak since an in-plane magnetic field is ineffective in setting the polarity of the N\'eel vector and therefore, the two opposite polarities causing an opposite sign of the anomalous Hall effect, are simultaneously present and keep the three-fold component small \cite{gonzalezbetancourt2023}.
Nevertheless, after subtraction of the even-in-field contribution from our ADMR data, residual signal displayed in Fig.~\ref{Figure4} is found. It shows that among the odd-in-field terms the three-fold contribution (due to the anomalous Hall effect) is dominant. When measured with a polarizing out of plane magnetic field the anomalous Hall effect strength can be, however, significantly increased \cite{gonzalezbetancourt2023}. In the following we will again focus on the even terms due to AMR, which dominate the magnetotransport for in-plane magnetic fields.

\begin{figure*}
        \centering
	\includegraphics[width=1.1\textwidth]{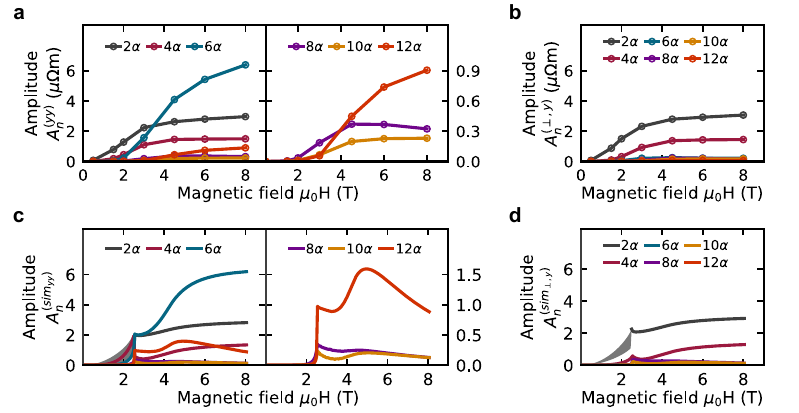}
	\caption{\textbf{Experimental and modeled magnetic field dependent resistivity components.} (a,c), and (b,d) show the variation of the amplitudes of the even order components of the longitudinal resistivity and transverse resistivity for current along $y$, respectively. Right panels in (a,c) show an enlarged view of the variation of the high order even contribution of the longitudinal resistivity. While panels (a,b) show experimental data, panels (c,d) show the data extracted from our modelling.}
	\label{Figure5}
\end{figure*}

We now use the Fourier analysis introduced above in order to determine the magnetic field dependence of the various harmonic components. We show in Fig.~\ref{Figure5}a,b (and supplementary Fig. S3) the field dependence of the amplitude of all even components of the longitudinal and transverse resistivities corresponding to the data shown in Fig.~\ref{Figure2} for current along $y$ ($x$). For both of these data sets we observe a gradual onset of the various amplitudes with a saturation tendency for fields above approximately 3~T. This suggests that above this amplitude the magnetic field is strong enough to induce a full rotation of the magnetic moments. 
Consistent with our previous analysis we find amplitudes and field dependence of the two-fold and four-fold contributions to correspond to each other when comparing longitudinal and transversal data. The six-fold contribution is detected only in the longitudinal resistivity and shows a somewhat distinct field dependence which either saturates only at higher fields or even has a contribution which does not saturate in the explored field range. In the right panel of Fig.~\ref{Figure5}a we focus on the higher order even components of the longitudinal resistivity. In particular, we find the twelve-fold component to have a field dependence similar to the one found for the six-fold one. It is, however, unclear if the twelve-fold component is indeed a distinct component or the result of a six-fold anisotropy contribution which causes the magnetic moments to deviate from the direction dictated by the magnetic field. 

\subsection{Stoner Wohlfarth modelling}

To obtain a better understanding of the field dependence and presence of different components, we set up a Stoner Wohlfarth model (\cite{stoner1948, ritzinger2023}) in which we assume a harmonic hexagonal anisotropy and limit the magnetic moments to the basal plane. For a given magnetic field direction and amplitude we minimize the energy with respect to the magnetic moment orientation and thus effectively obtain the history-dependent function $\theta$ of $\alpha$. Using this moment orientation and the AMR components determined by our symmetry analysis we mimic our magnetotransport data. We find that the magnetic moments remain nearly antiparallel and, under strong magnetic field, tend to align almost perpendicular to the applied magnetic field, regardless of the in-plane field direction. Therefore, we use solely the N\'eel vector orientation and ignore the small induced uncompensated moment in our analysis. The resulting calculated anisotropic resistivity data are then plugged into our Fourier component analysis. The field dependent amplitudes of these modelled data are shown in Fig.~\ref{Figure5}c,d. Note that we used the strength of the anisotropy and saturation amplitudes of the AMR components to obtain a field dependence similar to the experiments. Since the Stoner Wohlfarth model considers a single domain situation the modelling results are not unique for small fields. Depending on the relative orientation of the initially populated easy axis, a range of results can be obtained for magnetic fields below the spin-flop field. For higher field amplitudes, the moment's orientation is unique and does not depend on the starting conditions. The spin-flop transition causes a very sudden change in the model since no variation of the anisotropy barrier, domain wall pinning or thermal excitations are considered. It is furthermore noteworthy, that we plug into our model only AMR terms from Table~\ref{table1}, but we find also higher harmonic components in the Fourier analysis. In particular, we show in Fig.~\ref{Figure5}c,d the contribution with $n=8,10,$ and $12$ which arise in the simulation purely from the anisotropy and the AMR terms from Table \ref{table1}. Supplementary Fig.~S4 shows the detailed frequency and phase analysis of the modelling data which capture the same features as observed in the experiments. With our modelling we can describe the presence of all even harmonic components in the ADMR data by considering only the AMR terms from Table~\ref{table1}. However, the field dependence of the higher order even terms (in particular $n=12$) suggests that these components need to decay in the limit of very strong fields, which is when the anisotropy contribution to the energy becomes small compared to the Zeeman energy term.

\subsection{High field and temperature dependent magnetotransport}

\begin{figure*}
        \centering
	\includegraphics[width=1\textwidth]{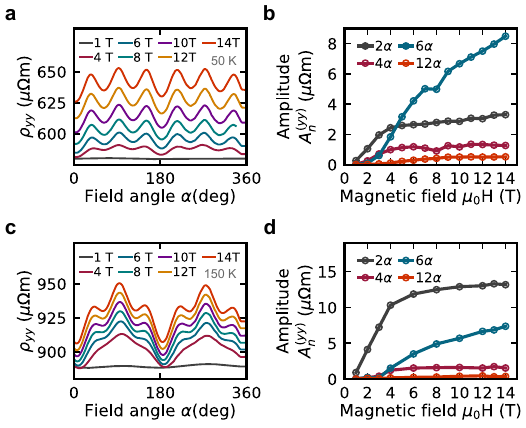}
	\caption{\textbf{Angular-dependent longitudinal magnetoresistance up to 14~T} 
    (a) Selected raw data for a Hall bar along $y$ direction, recorded at $T=50$~K. (b) Magnetic field dependence of the dominating frequency components at $T=50$~K. (c) Selected raw data for a Hall bar along $y$ direction, recorded at $T=150$~K.
    (d) Magnetic field dependence of the dominant frequency components at $T=150$~K. Note that data in (a,c) are shown as measured and no vertical offset was added between the curves.
 }
	\label{Figure6}
\end{figure*}

In order to clarify if such a decay can be found in experiments, we performed a second set of experiments. Here we focus on the variation of the longitudinal resistivity and used magnetic field up to 14~T. Figure~\ref{Figure6}a shows the measured data, while the resulting field dependent variation of the Fourier components is shown in Fig.~\ref{Figure6}b. Consistent with our previous experiments, we do find the terms with $n=2, 4, 6,$ and $12$ to dominate in the data. For this field range, our model suggests a visible decay of the twelve-fold component, however, no such decay, only a saturation behavior can be detected in the experiments. 

In Fig.~\ref{Figure6}a, in addition to the AMR also a field dependent isotropic magnetoresistance contribution can be observed. This can be extracted from our analysis as the term $n=0$ and is shown in supplementary Fig.~S5. We find an isotropic positive magnetoresistance of $\sim$10\% at a magnetic field of 14~T. Both the isotropic magnetoresistance and relative strength of the AMR components are changing with temperature. For ADMR measurements and analysis performed at a temperature of 150~K shown in Fig.~\ref{Figure6}c,d, we find the shape of the data is noticeably different. This is caused by a change in the weights of the various AMR components. In particular, we find that at increased temperature the two-fold non-crystalline AMR term dominates and the four and six-fold ones are relatively weaker. The magnetic field dependence of the non-crystalline AMR remains similar and saturates for high fields, while the six-fold term similar to the 50~K case shows no saturation.

\section{Discussion}

Our analysis of the angle dependent magnetotransport in altermagnetic MnTe reveals that there are various regimes of magnetic field strength which need to be considered somewhat separately. These are: (i) weak fields, i.e. below the spin flop transition, (ii) intermediate fields, i.e. just above the spin flop transition, (iii) strong fields aligning all moments along the field. In the weak fields the magnetic order is not fully rotated upon a magnetic field rotation, instead, it only slightly wiggles around the magnetic easy axis/axes. In this field regime, it is crucial to distinguish the angle of magnetic field in ADMR measurements from the angle of the magnetic moments governing the AMR. In this regime, also the population of multiple easy axes domains may play a crucial role. For compensated magnets, the field strength needed to overcome this regime (often called spin-flop field) can greatly vary. While in the canted antiferromagnet \ch{Sr2IrO4} a few 100~mT are sufficient \cite{fina2014, wang2019}, in near easy plane systems like \ch{Fe2As} \cite{wu2021} and MnTe several Tesla are required. In other systems like \ch{CrSb2}, no sign of saturation of the ADMR signal could be observed up to 24~T \cite{nakagawa2023}.
Once the characteristic field to rotate the compensated moments is overcome, i.e. for intermediate field strength, the magnetic moments can fully rotate coupled with the magnetic field. Nevertheless, a distinction of ADMR and AMR is still valid since the compensated moments tend to align nearly antiparallel and perpendicular to the magnetic field, merely with a minimal canting of the moments towards the magnetic field. In this regime, a near saturation of the ADMR variation can be observed in \ch{Fe2As} \cite{wu2021}. In MnTe, we also observe a saturation tendency, in particular of the harmonic components with two-, and four-fold symmetry. However, the component with six-fold symmetry continues to increase up to the maximum applied field of 14~T. It is conceivable that some ADMR components in compensated magnets saturate only under full saturation, i.e., for the highest magnetic fields, corresponding effectively to a parallel arrangement of the moments. Experimentally, this is typically reachable only for specially tuned anisotropies \cite{oh2023}, but otherwise, commonly beyond the experimentally available magnetic field strength. For increasing magnetic field, a continuous transition occurs between the canted rotation of the N\'eel order, regime (ii), and the full saturation, regime (iii). Often, in this regime, the magnetization induced by the magnetic field can not be ignored anymore. This might contribute to the lack of saturation of the six-fold ADMR component in MnTe. In our magnetoresistance modelling, a contribution of the induced magnetization resulting from the canting of the magnetic moments is ignored. Nevertheless, connected to the six-fold AMR combined with the six-fold anisotropy within the basal plane, our Stoner Wohlfarth model also reproduced the experimentally detected component with twelve-fold symmetry. Whether there is indeed such an intrinsic AMR component with twelve-fold symmetry present in MnTe, or our assumed Stoner-Wohlfarth model is too simple, is presently unclear. Future experiments at even higher fields might be able to clarify this. Theoretical efforts to explore the exact shape of the anisotropy, its microscopic origin \cite{vyborny2009}, and the impact of the moment orientation on the electronic structure are needed to develop a more precise model.

In conclusion, we have shown the angular dependence of the magnetoresistance in altermagnetic MnTe. While previous studies focused on the anomalous Hall effect or on magnetoresistance variation after field cooling, we show here that under an in-plane applied magnetic field, MnTe exhibits various anisotropic magnetoresistance contributions. In particular, we have identified both by theoretical symmetry analysis and magnetotransport experiments that harmonic variations of the resistance related to the in-plane orientation of the N\'eel vector are present in MnTe. Application of magnetic fields above $\sim 3$~T can induce a full rotation of the magnetic moments and experiments performed for different current directions allow us to confirm the functional dependence on current and moment direction.
At low temperature, the component with six-fold symmetry, related to the six-fold crystalline symmetry, dominates the magnetotransport. The more commonly observed non-crystalline AMR with two-fold symmetry becomes dominant at higher temperatures. In the experiments, we furthermore identified also higher order contributions which we could reproduce by model calculations. Their exact microscopic origin, however, remains elusive and more studies, at high magnetic fields and more first principle studies are needed. Altermagnetism manifests itself in the magnetotransport as a three-fold contribution to the transverse magnetoresistance due to the anomalous Hall effect.
Our work contributes to the understanding of the magnetotransport properties of MnTe which is evolving as a workhorse material for altermagnetic spintronics.

\section{Methods}

\subsection{Symmetry analysis of the resistivity tensor}

To analyze the symmetry of the resistivity tensor $\rho_{ij}$, we expand it in the N\'eel vector $\mathbf{L}$:

\begin{align}
    \rho_{ij}(\mathbf{L}) = \rho_{ij}^{(0)} + \rho_{ijk}^{(1)} L_k + \rho_{ijkl}^{(2)} L_l L_l + \dots
\end{align}

We determine the symmetry of each term in this expansion, using the code Symmetr \cite{zelezny}. Since the anisotropy of our MnTe limits $\mathbf{L}$ to the $ab$ plane we use $\mathbf{L}/L = (\cos(\theta),\sin(\theta),0)$, where $\theta$ denotes the angle between the $x$-axis and $\mathbf{L}$. Substituting this for $\mathbf{L}$ we obtain an expansion in powers of $\cos(\theta)$ and $\sin(\theta)$. This can be converted to expansion in $\cos(n\theta)$ and $\sin(n\theta)$, where $n$ are integers. Note that in general terms of $n$-th order in the powers of $\cos(\theta)$ and $\sin(\theta)$ expansion correspond to $n$-th and \textit{lower} order terms in the $\cos(n\theta)$ and $\sin(n\theta)$ expansion.  Writing down the current as $\mathbf{j} = (\cos(\varphi),\sin(\varphi),0)$, considering terms up to sixth order, and separating the components parallel and perpendicular to current, we obtain the results in Table \ref{table1}.

\subsection{Thin film growth}

The single crystalline epitaxial $\alpha$-MnTe thin films are grown by molecular beam epitaxy (MBE) on InP (111)A substrates. A two-dimensional growth of MnTe, as judged by streaked RHEED pattern, is achieved at substrate temperatures in the range of 370$^{\circ}$C-450$^{\circ}$C. The samples studied here have a thickness of 48~nm and 35~nm. For structural characterization we carried out X-ray diffraction radial scans as well as polefigure measurements with CuK$\alpha_1$ radiation. The orientation of our layers on the substrate is $(0001)[100]_{\rm MnTe} || (111)[11\bar2]_{\rm InP}$, i.e. $c$-axis is the out-of-plane direction and the hexagonal basel plane is inside the sample plane.

\subsection{Lithographic processing of Hall bars}

Hall bars were fabricated in different orientations on our \ch{MnTe} thin films. For the patterning we used electron beam lithography, argon-based ion beam etching or argon ion milling in a plasma etcher, and electron beam evaporation of Cr/Au contacts. Data in Fig.~\ref{Figure2}a,c are from two different Hall bars along the $x$ direction ($[2\bar{1}\bar{1}0]$) with width of 50 $\mu$m and longitudinal contacts spaced 500 $\mu$m apart, and width of 30 $\mu$m and longitudinal contacts spaced 300 $\mu$m apart, respectively. Data in
Fig.~\ref{Figure2}b,d and Fig.~\ref{Figure6} are from two separate Hall bars oriented along the $y$ direction ($[01\bar{1}0]$) with a width of 30 $\mu$m, and the longitudinal contacts are positioned 300 $\mu$m apart. Thin films from multiple growth runs were used in our experiments. 

\subsection{Measurement setup}
The magneto-transport measurements were performed using two different setups. 
The first setup utilizes a Quantum Design Physical Property Measurement System (PPMS) with in-built electrical transport option (ETO) and sample holder (PCB 7084) allowing for in-plane field rotations. It comprises a precision current source and voltage pre-amplifiers coupled to a digital signal processing (DSP) unit. A sinusoidal AC drive current is applied. The DSP unit is then used to record the voltage response and filter the portion of the signal at the same frequency as the drive current. Typical frequency and current amplitude used were 18.3~Hz and 0.01~mA, respectively, which corresponds to approximately $5\times10^6$ A/m$^2$.

On the second setup, a higher magnetic field capacity of up to 14~T was used. Although the same measurement procedure was followed, an alternating current using a Keithley 6221 AC current source was applied and a Zurich MFLI lock-in amplifier 500 KHz - 5 MHz was used to record the data. In both cases samples were electrical contacted by wedge bonding and BNC and RCA electrical connectors were used. The sample temperature was sufficiently stabilized before the ADMR data were recorded by a step-wise rotation of the sample in a constant magnetic field, i.e. during the data acquisition no movement or field sweeps were performed. The samples were rotated both in clockwise and anticlockwise fashion.

For magnetotransport measurements with current along the $x$-axis, the contact configuration depicted in Fig.~1d was rotated by 90 deg. This includes the polarity of the transversal voltage detection which leads to the fact that $\rho_{\perp,x}$ corresponds to $\rho_{yx}$, and $\rho_{\perp,y}$ to $\rho_{\bar{x}y}$, where $\bar x$ indicates the change in the polarity. Employing this nomenclature allows for a direct comparison with data in Table 1.

\subsection{Fourier component analysis}

To analyze our data we perform a discrete Fourier transform of the angular-dependent magnetoresistance data. The resulting complex frequency components are converted to the amplitude and phase of relevant harmonics. This means we effectively decompose our signal as follows:

\begin{equation}
\rho_{\parallel} = \sum_{n=0} A_n \cos(n\alpha + \alpha_n)
\end{equation}

where $A_n$ represents the amplitude of the $n$-th harmonic, $\alpha$ represents the angle at which the magnetic field is applied with respect to the $x$ axis $([2\bar{1}\bar10])$, and $\alpha_n$ the phase offset of the $n$-th harmonic.
We analyze the amplitude and phase spectrum and focus on the initial $15$ harmonics since we find components with $n>15$ to be negligible. 

\subsection{Stoner Wohlfarth model of MnTe}

\renewcommand{\vec}[1]{\mathbf{#1}}

The equilibrium orientation of magnetic moments in MnTe was determined by considering a six-fold magnetocrystalline anisotropy, exchange energy, and Zeeman energy within a single-domain model. For the total energy $E$ per sample volume $V$ we use:
\begin{equation}
E/V = J_{\rm ex} \vec{\hat M}_1 \cdot \vec{\hat M}_2 - \vec B \cdot (\vec M_1 + \vec M_2) + E_{\rm MAE}(\vec{\hat M}_1) + E_{\rm MAE}({\vec{\hat M}}_2)
\end{equation}
where $J_{\rm ex}$ is the exchange constant, $E_{\rm MAE}$ is the magneto-crystalline anisotropy function, $\vec{M}_i$ are the sublattice magnetizations, and $\vec{B}$ the magnetic field. A hat (''\ $\hat{}$\ '') denotes a unit vector.
We restrict ourselves to a 2D description within the basal plane, which is sufficient to describe our in-plane magnetic field rotations. Within the basal plane we use the easy axis orientation (one of the $<01\bar{1}0>$ directions) as our reference direction and define the angle of the magnetic field $\alpha$ and the angle of the magnetic moments  ($\psi_1, \psi_2$) of the sublattice with respect to this axis.
Inserting the angular positions of the sublattices, magnetic field, and the exchange field defined as $B_{\rm ex} = J_{\rm ex}/M$ with $M = \left|\vec M_{1,2}\right|$ this can be rewritten as
\begin{align}
E/V =& - M B_{\rm ex} \cos( \psi_1 - \psi_2 ) - M B \left[ \cos\left(\alpha - \psi_1\right) + \cos\left(\alpha -\psi_2\right) \right] \nonumber \\
     & + E_{\rm MAE}(\psi_1) + E_{\rm MAE}(\psi_2).
\label{eq:energy}
\end{align}
The exchange field is estimated from the N\'eel temperature as $B_{\rm ex} = k_{\rm B}T_{\rm N}/\mu_{\rm B}$, where we use the Boltzmann constant $k_{\rm B}$ and the Bohr magneton $\mu_{\rm B}$. The magneto-crystalline anisotropy energy density is expressed as
\begin{equation}
    E_{\rm MAE}(\psi) = K_{\rm MAE} \sin^2 3\psi,
\end{equation}
following from the six-fold in-plane symmetry of the hexagonal material. Here, $K_{\rm MAE}$ has units of energy density and is used as a free parameter. It is chosen such that a resulting spin-flop field of $\sim2.5$ T is obtained in the model.
From a numerical minimization using the conjugate gradient method we obtain the orientations of the magnetic moments ($\psi_1, \psi_2$). Using the resulting N\'eel vector orientation, given by the angle $\varphi = (\psi_1 + \psi_2 -\pi)/2$, we calculate the contribution to the anisotropic magnetoresistance as given in Tab.~\ref{table1}.

\section{Competing interests}

All authors declare no financial or non-financial competing interests. 

\section{Data Availability}

All data are available in the main text or the supplementary materials. Further data are available from the corresponding author on reasonable request.

\section{Code Availability}

In the manuscript, we used the code Symmetr to analyze the symmetry of the magnetotransport response Mozilla Public License. Symmetr is published under the Mozilla Public License at \url{https://bitbucket.org/zeleznyj/linear-response-symmetry}. All input files are available from the authors on request.

\section{Author contributions}

R.D.G.B. and D.K. conceived the idea, proposed the experiments and co-wrote the initial manuscript.
G.S. prepared the thin film samples.
R.D.G.B., J.Z. and K.G. fabricated devices and performed magnetotransport measurements.
J.\v{Z}. performed the symmetry analysis. B.R., H.R. characterized the samples.
D.K., P.R. and K.V. performed the model calculations.
T.K. and R.D.G.B. performed high field measurements.
D.K., T.J., K.O., A.T. and B.B. supervised and analyzed data.
All authors contributed to data analysis and read and commented on the manuscript.

\textbf{Acknowledgements}

We acknowledge the support by the Czech Science Foundation (Grant No. 22-22000M) as well as Lumina Quaeruntur fellowship LQ100102201 of the Czech Academy of Sciences. Experiments were performed in MGML (mgml.eu), which is supported within the program of Czech Research Infrastructures (project no. LM2023065).
This work was supported by HLD-HZDR, member of the European Magnetic Field Laboratory (EMFL).
We would like to express our gratitude to M. Uhlarz for his assistance in setting up the experiment.
Support by the Austrian Science Funds, (Grant No. AI0656811/21) and the LIT Grant No. LIT-2022-11-SEE-131 of the University of Linz is also acknowledged. 
The study was also supported by Charles University, project GA UK No. 266723.

%

\end{document}